\documentclass{epl}
\usepackage{lscape,graphicx}

\title{Microscopic theory for the glass transition in a system
without static correlations}
\author{R. Schilling\inst{1} \and G. Szamel\inst{1,2}}
\institute{
  \inst{1} Institut f\"ur Physik, Johannes Gutenberg-Universit\"at
Mainz, D-55099 Mainz, Staudinger Weg 7, Germany\\
  \inst{2} Department of Chemistry, Colorado State University,
Ft. Collins, CO 80523, USA }

\begin{document}

\maketitle

\begin{abstract}
We study the orientational dynamics of infinitely thin hard rods of 
length
$L$, with the centers-of-mass
fixed on a simple cubic lattice with lattice constant $a$.
We approximate the influence
of the surrounding rods onto dynamics of a pair of rods by
introducing an effective rotational diffusion constant $D(l)$, 
$l=L/a$.
We get $D(l) \propto [1-\upsilon(l)]$, where
$\upsilon(l)$ is given through an integral
of a time-dependent torque-torque correlator of an isolated pair
of rods. A glass transition occurs at $l_c$, if $\upsilon(l_c)=1$. We
present a variational and a numerically exact evaluation of 
$\upsilon(l)$.
Close to $l_c$ the diffusion constant decreases as $D(l) \propto
(l_c-l)^\gamma$, with $\gamma=1$. Our approach predicts a glass
transition in the absence of any static correlations, in contrast
to present form of mode coupling theory.
\end{abstract}

When a system of interacting species is cooled sufficiently fast
it will typically undergo a glass transition. In the last two
decades there has been a significant progress in microscopic
understanding of this transition. The first approach is the
so-called mode coupling theory (MCT) \cite{1,2}, first suggested
and mainly worked out by G\"otze and coworkers. For reviews see
Refs. \cite{3,4,5}. MCT yields a closed set of equations for
time-dependent correlators. This set involves as an input
\textit{static} correlators which depend only on thermodynamic
variables like temperature $T$, density $n$, etc. With decreasing
$T$ (or increasing $n$) local ordering described by peaks of the
static correlators grows. When the peaks' magnitude becomes large
enough MCT predicts an ergodicity breaking transition at a
critical temperature $T_c$ (or density $n_c$). This
\textit{dynamic} transition is interpreted as an ideal glass
transition.

A different approach has recently been proposed by M\'ezard and
Parisi\cite{6,7}. Inspired by the replica theory for spin glasses
they developed a microscopic replica theory for structural
glasses. This theory predicts a \textit{static} glass transition
at a temperature $T_f$ which is lower than $T_c$. One of its
characteristic features of this transition is the vanishing of the
configurational entropy at $T_f$. The physical picture behind that
replica theory was suggested earlier by Singh \textit{et al}.
\cite{8}.

\begin{figure}
\includegraphics[height=7.8cm,angle=270]{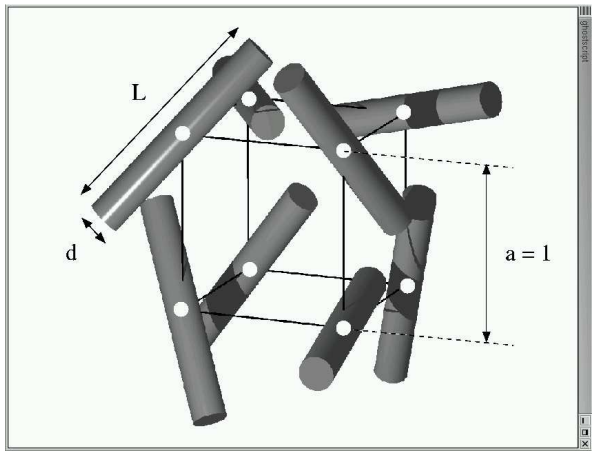}
\caption{\label{fig1}
Unit cell of the cubic lattice with hard rods of
length $L$ and diameter $d$ with their centers at the lattice sites.}
\end{figure}

Despite providing new valuable insights, these microscopic
theories can \textit{not} be complete, because there exists a
\textit{purely dynamical} mechanism which leads to a glass
transition which is \textit{not} based on the existence of any
\textit{static} correlations. As an example, let us consider a
simple cubic lattice with $N$ lattice sites and lattice constant
$a$ with hard rods of length $L$ and diameter $d$ with their
centers fixed on the lattice sites (Fig.~\ref{fig1}). This model
system was first introduced by Rubinstein and Obukhov \cite{13}.
It can be used to describe orientational glasses\cite{9,10}
\footnote{We stress that this type of model is quite different to
the mixed crystals studies by Michel \cite{11} where the glass
transition occurs due to \textit{quenched disorder}}. The increase
of the reduced rod length $l=L/a$ leads to an increase of the
\textit{steric hindrance}, which may lead to a glass transition.
Now suppose that the thickness $d$ goes to zero, for fixed $a$ and
$L$. For $d=0$ all static orientational correlations vanish and
there is no static phase transition. Nevertheless, the
simulational studies by Renner \textit{et al}. \cite{14} for a
fcc lattice and Obukhov \textit{et al}. \cite{15} clearly
demonstrate the existence of a (dynamical) glass transition at a
critical length $l_c=L_c/a$. In addition, it is particularly
interesting that the time-dependent orientational correlators
studied in Ref. \cite{14} exhibit a two-step relaxation process
similar to that observed for supercooled liquids and predicted by
MCT\cite{3,4,5}. However, as Renner \textit{et al}. \cite{14}
stressed the MCT in its present form fails to describe this
dynamical transition. The reason is obvious: the vertices which
enter into the MCT equations vanish\footnote{There exist
non-generic model systems for which the vertices do not vanish
although there are no static correlations \cite{Fuchs}.}. Thus,
\textit{present} formulation of MCT as well as the replica theory
\cite{6,7} cannot explain the numerical findings of Ref.
\cite{14,15}, due to the complete trivial statics for $d=0$.

This discussion demonstrates the necessity to derive a microscopic
theory which is capable to describe a purely dynamical glass
transition for systems with completely trivial statics. To achieve
this for the infinitely thin hard rods on a cubic lattice is the
goal of the present contribution. Details and additional features
of our approach will be given elsewhere \cite{16}.

Let us consider the time-dependent 1-rod density $\rho_{n_1}^{(1)}
(\Omega; t)$ for a tagged rod at lattice site $n_1$ with
orientation $\Omega=(\Theta, \phi)$ and
\begin{equation} \label{eq1}
\rho^{(1)}_{n_1} (\Omega; 0) = \delta (\Omega | \Omega')
\end{equation}
as initial condition, where $\delta (\Omega|\Omega')=\sin \Theta
\, \delta (\Theta-\Theta') \, \delta (\phi - \phi')$. If $l < 1$
the rods cannot touch and the tagged rod will perform a free
\textit{rotational} motion. However, for $l > 1$ steric hindrance
exists and the tagged rod's motion becomes influenced by its
nearest, next-nearest, etc. neighbours. These interactions will
influence the rotational motion of the rod at site $n_1$. With
increasing $l$ rotational diffusion constants will decrease and
finally may become zero at a critical length $l_c$.

The time derivative of $\rho^{(1)}_{n_1} (\Omega; t)$ is related
to the 1-rod current density ${\bf j}_{n_1}^{(1)} (\Omega;t)$ by
the continuity equation:
\begin{equation} \label{eq2}
\frac{\partial} {\partial t} \, \rho_{n_1}^{(1)} \, (\Omega; t)
+ {\bf \nabla}_{\Omega} \cdot {\bf j}^{(1)}_{n_1} \, (\Omega; t) =0.
\end{equation}
We define a generalized rotational diffusion
tensor, $\tens{D}(\Omega,\Omega'; t)$, by:
\begin{equation} \label{eq3}
{\bf j}^{(1)}_{n_1} (\Omega; t) =- \int\limits_0^t \, dt' \,
\int d \Omega' \,\, \tens{D} (\Omega, \Omega';
t-t') \cdot {\bf \nabla}_{\Omega'} \rho_{n_1}^{(1)} (\Omega'; t').
\end{equation}

\begin{figure}
\includegraphics[height=7.8cm,angle=270]{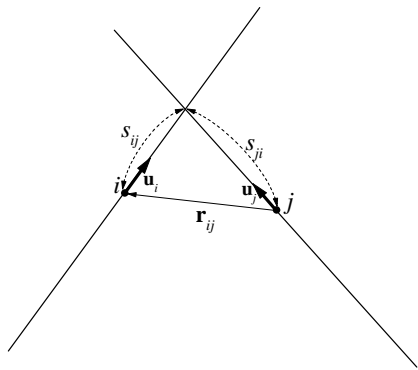}
\caption{\label{fig2}
Illustration of the geometrical quantities defined in the text for
two needles $i$ and $j$.}
\end{figure}

In the following we derive a closed equation for the diffusion
tensor $\tens{D}$ which will allow to investigate its
$l$-dependence. To derive this equation we use Smoluchowski
(Brownian) dynamics (glass transition should not depend on the
microscopic dynamics \cite{17,18}). We start from the so-called
generalized Smoluchowski equation for the $N$-rod probability
density $P_N (\Omega_1, \cdots, \Omega_N; t)$:
\begin{equation} \label{eq4}
\frac{\partial} {\partial t} P_N (\Omega_1, \cdots, \Omega_N; t)
= D_0 \sum\limits_{n=1}^N \, {\bf \nabla}_n \cdot
[{\bf \nabla}_n - {\bf
T}_n (\Omega_1, \cdots , \Omega_N) ] P_N (\Omega_1, \cdots,
\Omega_N; t)
\end{equation}
where $D_0$ is the \textit{bare} rotational diffusion constant,
${\bf \nabla}_{n_1} \equiv {\bf \nabla}_{\Omega_n}$, and
\begin{equation} \label{eq5}
{\bf T}_n (\Omega_1, \cdots , \Omega_N)= \sum\limits_{j \neq n} \,
\, {\bf T}_{jn} (\Omega_j, \Omega_n)
\end{equation}
where
\begin{equation} \label{eq6}
{\bf T}_{ij} (\Omega_i, \Omega_j) =s_{ij} ({\bf u}_i \times {\bf
\hat{r}}_{ij}^\bot) \, \delta (r_{ij}^\bot - 0^+) \Theta
\left(\frac{L}{2}- |s_{ij}|\right) \Theta \left(\frac{L}{2}
-|s_{ji}|\right)
\end{equation}
describes the singular torque when two rods at sites $i$ and $j$
with vector distance ${\bf r}_{ij}$
touch each other \cite{19} \footnote{Since ${\bf \nabla}_{u_n}$ is
used in Ref. \cite{19} instead of ${\bf \nabla}_{\Omega_n}$ their
expression for ${\bf T}_{ij}$ differs slightly from ours.}. ${\bf
r}_{ij}^\bot$ is the component of ${\bf r}_{ij}$ perpendicular to
the plane defined by the unit vectors along the rods, ${\bf u}_i$
and ${\bf u}_j$, $r_{ij}^\bot=| {\bf r}_{ij} ^\bot|$ and  $ {\bf
\widehat{r}}_{ij}={\bf r}_{ij}^{\bot} /r_{ij}.$ $s_{ij}$ and
$s_{ji}$ (which depend on $\Omega_i$, $\Omega_j$) is the distance
of the contact point to the center of rod $i$ and rod $j$,
respectively (cf. Fig.~\ref{fig2}). $\delta(x-0^+)$ stands for
$\lim\limits_{d \rightarrow0^+} \, \delta (x -d).$ The reader
should note that ${\bf r}_{ij}$ is t-independent and is equal to
the lattice vector ${\bf R}_{ij}={\bf R}_i- {\bf R}_j$, in
contrast to a \textit{fluid} of hard rods (needles).

A generalized Smoluchowski equation similar to Eq. (\ref{eq4}) had
been used to describe a \textit{fluid} of infinitely thin rods
with randomly frozen orientations \cite{20,21}. In this case there
is no glass transition if the longitudinal component
of the bare translational diffusion tensor is
positive.

From Eq.~(\ref{eq4}) one can derive a hierarchy of equations of
motion for the reduced $j$-rod densities $\rho^{(j)}_{n_1 \cdots
n_j} (\Omega_1, \cdots, \Omega_j; t)$ \cite{22}.
For instance, for $j=1$ one gets:
\begin{equation} \label{eq7}
\frac{\partial}{\partial t}\, \rho_{n_1}^{(1)} (\Omega_1;t) = D_0
{\bf \nabla}_{n_1} \cdot \left[{\bf
 \nabla}_{n_1} \rho_{n_1}^{(1)} (\Omega_1; t) -
\sum\limits_{n_2} \int d \Omega_2 \, {\bf T}_{n_1 n_2} (\Omega_1 ,
\Omega_2) \rho^{(2)}_{n_1 n_2} (\Omega_1, \Omega_2; t)\right].
\end{equation}

The first term on the r.h.s. describes the free Brownian motion.
The second term describes the influence of a rod at site $n_2$ on the
tagged rod. It involves the 2-rod density
$\rho^{(2)}_{n_1 n_2} (\Omega_1, \Omega_2; t)$. To proceed it is
convenient to introduce the \textit{fluctuations}:
\begin{equation} \label{eq8}
\delta \rho^{(j)}_{n_1 \cdots n_j} (\Omega_1, \cdots, \Omega _j;
t)= \rho_{n_1 \cdots n_j}^{(j)} (\Omega_1, \cdots , \Omega_j; t)-
\frac{1} {(4 \pi)^{j-1}} g^{(j)}_{n_1 \cdots {n_j}} (\Omega_1,
\cdots , \Omega_j)
 \rho^{(1)}_{n_1}(\Omega_1;t)
\end{equation}
where $g_{n_1 \cdots n_j}^{(j)} (\Omega_1, \cdots, \Omega_j)$ is
the equilibrium $j$-rod distribution function. Note that the
equilibrium distribution functions are equal to one almost
everywhere.

Using Eq.~(\ref{eq8}) one obtains a hierarchy of equations for
$\delta \rho^{(j)}$. The lowest level of this hierarchy is still
given by Eq.~(\ref{eq7}) where one is allowed to replace
$\rho^{(2)}_{n_1 n_2}$ by $\delta \rho^{(2)}_{n_1 n_2}$, because
$\sum\limits_{n_2} \,\, \int d \Omega_2 \,\,\,{\bf T}_{n_1 n_2}
\,(\Omega_1, \Omega_2)  \,\, g^{(2)}_{n_1 n_2} (\Omega_1,
\Omega_2) =0$. At the second level one gets \cite{16}:
\begin{eqnarray} \label{eq9}
&&\frac{\partial}{\partial t} \, \delta\rho^{(2)}_{n_1 n_2}
(\Omega_1, \Omega_2; t) = - \frac{1} { 4 \pi} \big( {\bf
\nabla}_{n_1} \,\, g_{n_1 n_2}^{(2)} \, (\Omega_1, \Omega_2) \big)
\, \cdot
{\bf j}^{(1)}_{n_1} (\Omega_1; t) + \nonumber \\
&&D_0 \left\{ {\bf \nabla}_{n_1} \cdot\big[{\bf \nabla}_{n_1} - {\bf
T}_{n_1 n_2} \big(\Omega_1, \Omega_2 \big) \big] + \big( 1
\leftrightarrow 2 \big) \right\} \, \delta \rho^{(2)}_{n_1 n_2}
(\Omega_1, \Omega_2; t) + A^{(2)}_{n_1 n_2} (\Omega_1, \Omega_2;
t)
\end{eqnarray}
where
\begin{eqnarray} \label{eq10}
A^{(2)}_{n_1 n_2} (\Omega_1, \Omega_2; t) &=& D_0 {\bf
\nabla}_{n_1}\cdot\left[ g^{(2)}_{n_1 n_2} (\Omega_1, \Omega_2)
\sum\limits_{n_3} \frac{1}{4 \pi} \int \, d \Omega_3 {\bf T}_{n_1
n_3} (\Omega_1, \Omega_3) \delta \rho^{(2)}_{n_1 n_3}
(\Omega_1, \Omega_3; t) \right] \nonumber\\
&&- D_0 \, \sum\limits_{\nu=1}^{2}  {\bf \nabla}_{n_\nu} \cdot
\sum\limits_{n_3} \int \, d \Omega_3 {\bf T}_{n_\nu n_3}
(\Omega_\nu, \Omega_3) \delta\rho^{(3)}_{n_1 n_2 n_3} (\Omega_1,
\Omega_2, \Omega_3; t) .
\end{eqnarray}
$A^{(2)}_{n_1 n_2}$ describes the influence of a third rod at
$n_3$ on the rods at $n_1$ and $n_2$. Here we follow Refs.
\cite{20,21} and approximate this influence by introducing an
effective diffusion tensor $\tens{D}^{\mathrm{eff}}$.
Specifically, in Eq. (\ref{eq9}) we replace $D_0$ by the
effective diffusion tensor and we neglect $A^{(2)}_{n_1 n_2}$:
\begin{eqnarray} \label{eq11}
&& \left\{ D_0 {\bf \nabla}_{n_1} \cdot
\left[{\bf \nabla}_{n_1} - {\bf
T}_{n_1 n_2} (\Omega_1, \Omega_2) \right] + \big( 1
\leftrightarrow 2 \big) \right\} \delta \rho^{(2)}_{n_1 n_2}
(\Omega_1, \Omega_2; t) \rightarrow \\
&& \left\{ {\bf \nabla}_{n_1}\cdot\int\limits_0^t \, dt' \int \, d
\Omega_1' \tens{D}^{\mathrm{eff}} (\Omega_1, \Omega_1'; t-t') \cdot
\left[ {\bf \nabla}_{n_1'} - {\bf T}_{n_1 n_2} ( \Omega_1',
\Omega_2) \right] + \big( 1\leftrightarrow 2 \big) \right\} \delta
\rho^{(2)}_{n_1 n_2}(\Omega_1', \Omega_2;t') \nonumber
\\ \label{eq11a}
&& A^{(2)}_{n_1 n_2} (\Omega_1, \Omega_2; t) \approx 0 .
\end{eqnarray}
In order to get a self consistent equation we then set in
Eq.~(\ref{eq11}) $\tens{D}^{\mathrm{eff}} \equiv  \tens{D}$. This
approximation leads to a closed set of equations, given by
Eqs.~(\ref{eq3}),~(\ref{eq9}),~(\ref{eq11}) and ~(\ref{eq11a}).
Taking Laplace transforms of these equations one finally gets a
self consistent equation for the diffusion tensor
$\tens{D}(\Omega_1, \Omega_2)= \lim\limits_{z \rightarrow 0}
\tens{D}(\Omega_1, \Omega_2; z )$, where the latter is the Laplace
transform of $\tens{D}(\Omega_1, \Omega_2; t)$\cite{16}. For
\textit{technical} convenience we perform here an additional
approximation:
\begin{equation} \label{eq12}
D^{\alpha \beta}(\Omega_1, \Omega_1') \approx D (l)\delta^{ \alpha
\beta} \delta (\Omega_1|\Omega_1').
\end{equation}
Substitution of Eq.~(\ref{eq12}) into the self consistent equation
for $\tens{D}$ yields \cite{16}:
\begin{equation} \label{eq13a}
D(l) = D_0 [1 - \upsilon(l)]
\end{equation}
with the $l$-dependent coupling function
\begin{eqnarray} \label{eq13}
\upsilon(l) = \frac{1}{3}
\sum\limits_{n_2} \int\limits_0^\infty \, dt \langle
{\bf T}_{n_1 n_2} (\Omega_1, \Omega_2)\cdot e^{{\cal L}^{(2)\dagger}t}
\, {\bf T}_{n_1 n_2} (\Omega_1, \Omega_2) \rangle \quad.
\end{eqnarray}
where ${\cal L}^{(2) \dagger}$ is an adjoint Smoluchowski operator
for an \textit{isolated} pair of rods
\begin{equation} \label{eq14}
{\cal L}^{(2) \dagger} = [{\bf \nabla}_{n_1} + {\bf T}_{n_1 n_2}
(\Omega_1, \Omega _2)]\cdot{\bf \nabla}_{n_1} +
[ {\bf \nabla}_{n_2} + {\bf T}_{n_2 n_1} (\Omega_2,\Omega_1) ]
\cdot {\bf \nabla}_{n_2}.
\end{equation}
The time-dependent torque-torque correlator on the r.h.s. of
Eq.~(\ref{eq13}) is defined by:
\begin{eqnarray} \label{eq15}
&&\langle {\bf T}_{n_1 n_2} (\Omega_1, \Omega_2 ) \cdot
e^{{\cal L}^{(2)\dagger} t} \, \, {\bf T}_{n_1 n_2} (\Omega_1,
\Omega_2) \rangle \nonumber\\
&& =\frac{1} {(4 \pi)^2} \,\, \int d \Omega_1 \, \int d \Omega_2
\,\, g^{(2)}_{n_1 n_2} (\Omega_1, \Omega_2) \, {\bf T}_{n_1
n_2}(\Omega_1, \Omega_2)\cdot e^{{\cal L}^{(2) \dagger}t} \, {\bf
T}_{n_1 n_2} (\Omega_1, \Omega_2) .
\end{eqnarray}
Due to the lattice translation invariance $\upsilon(l)$ does not
depend on $n_1$. Moreover, $\upsilon(l) \geq 0$, where the
equality sign holds only for $l \leq 1$ and $\upsilon(l) \sim l^3$
for $l \gg 1$.

The coupling constant $\upsilon(l)$ can be calculated via a
Brownian dynamics simulation of the isolated two-rod system or it
can be estimated analytically. The former approach is
straightforward. In the following we briefly discuss the latter.
Eq. (\ref{eq13}) can be rewritten as
\begin{equation} \label{eq16}
\upsilon(l)= \frac{1}{3}
\sum\limits_{n_2} \, \langle {\bf T}_{n_1 n_2}(\Omega_1, \Omega_2)
\cdot (-{\cal L}^{(2) \dagger})^{-1} \, {\bf T}_{n_1 n_2}
(\Omega_1, \Omega_2) \rangle.
\end{equation}
We introduce a vector function ${\bf f}_{n_1 n_2} (\Omega_1,
\Omega_2)$ such that
\begin{equation} \label{eq17}
{\cal L}^{(2) \dagger}
\, \, {\bf f}_{n_1 n_2} (\Omega_1, \Omega_2) = {\bf
T}_{n_1 n_2} (\Omega_1, \Omega_2) \quad .
\end{equation}
If ${\bf f}$ is a solution of Eq.~(\ref{eq17}), then $\upsilon(l)$
can be readily calculated: $\upsilon(l)= -\frac{1}{3}
\sum\limits_{n_2} \langle {\bf T}_{n_1 n_2} (\Omega_1, \Omega_2)
\cdot {\bf f}_{n_1 n_2}
(\Omega_1, \Omega_2)\rangle$. ${\bf f}_{n_1 n_2} (\Omega_1,
\Omega_2)$ can be determined from the variational equation of the
following functional (for a similar discussion see Ref. \cite{23}):
\begin{eqnarray} \label{eq18}
&&\mathcal{F} [ \{{\bf f}_{n_1 n_2} (\Omega_1, \Omega_2)
\}]\nonumber\\
&&= \frac{1}{3}
\langle {\bf f}_{n_1 n_2} (\Omega_1, \Omega_2) \cdot {\cal L}^{(2)
\dagger}{\bf f}_{n_1 n_2} (\Omega_1, \Omega_2) \rangle - \frac{2}{3}
\langle {\bf f}_{n_1 n_2} (\Omega_1, \Omega_2)\cdot{\bf T}_{n_1 n_2}
(\Omega_1, \Omega_2) \rangle.
\end{eqnarray}
$\upsilon(l)$ is the maximum value of $\mathcal{F}$ and
$\upsilon^{\mathrm{var}}(l)\equiv \mathcal{F} [\{ {\bf f}_{n_1
n_2}^{\mathrm{var}} \, (\Omega_1, \Omega_2) \}] \leq \,
\upsilon(l)$ for all trial functions ${\bf f}_{n_1
n_2}^{\mathrm{var}}$. We have used the following simple trial
function
\begin{eqnarray} \label{eq19}
{\bf f}_{n_1 n_2}^{\mathrm{var}}(\Omega_1, \Omega_2)=\lambda
\sin\Theta_1 \sin \Theta_2 \left (
\begin{array} {*{3}c@{\;}} \cos \Theta_1  \cos \phi_1 \\ \cos
\Theta_1
\sin \phi_1 \\ -\sin\Theta_1
\end{array} \right ) \cdot \left \{
\begin{array}{ll} \frac{\pi}{2} -(\phi_2-\phi_1),
& 0 < \phi_2-\phi_1 <\pi
\\
\frac{3 \pi}{2} - (\phi_2- \phi_1), & \pi < \phi_2-\phi_1 < 2 \pi
\end{array} \right.
\end{eqnarray}
where $\lambda$ is a variational parameter. Note that ${\bf
f}^{\mathrm{var}}_{n_1 n_2}$ is independent on $n_1$, $n_2$ and
does not depend explicitly on $l$. For this simple trial function
the $l$-dependence of $\upsilon^{\mathrm{var}}(l)$ comes from that
of ${\bf T}_{n_1 n_2} (\Omega_1, \Omega_2)$.

\begin{figure}
\includegraphics[height=7.8cm,angle=270]{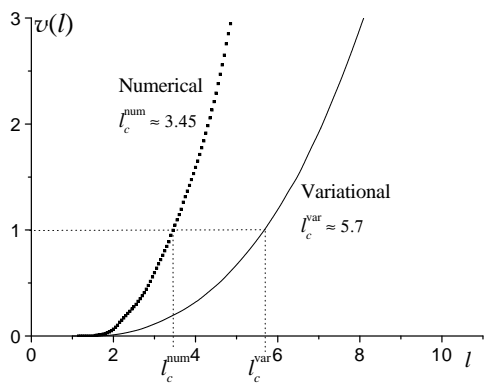}
\caption{\label{fig3}
$\upsilon(l)$ for a simple cubic lattice
from the numerically exact (squares) and the
variational calculation (solid line). $l_c$ denotes the critical
length for which $\upsilon(l_c)=1$.}
\end{figure}

In Fig.~\ref{fig3} we present $\upsilon(l)$ for a simple cubic lattice 
as obtained from Eq.~(\ref{eq13}) by a computer simulation of the
isolated 2-rod system and from the variational approach. The glass
transition is defined by $D(l_c)=0$ which results in
$\upsilon(l_c)=1$. A finite value for $l_c > 1$ must exist because
$\upsilon(l)\ge 0$ and $\upsilon(l) \sim l^3$ for $l \gg 1$. From
Fig.~\ref{fig3} we get $l_c^{\mathrm{num}} \cong 3.45$ and $l_c
^{\mathrm{var}} \cong 5.7$, respectively. For the fcc lattice we
have found \cite{16} $l_c^{\mathrm{num}} \cong 2.2 $ and
$l_c^{\mathrm{var}} \cong 3.96$ which reasonably fits
$l_c^{\mathrm{MD}} \cong 2.7$ \footnote{Obukhov \textit{et al.}
\cite{15} considered a system of rods attached to the lattice
sites of an sc lattice at one of their end points and obtained a
slightly different value, $l_c \cong 4.5$.} from the MD-simulation
\cite{14}. From this point of view our microscopic approach,
including all approximations, leads to a satisfactory value for
the critical length. Ref. \cite{14} has also demonstrated that the
rotational diffusion constant follows a power law $D(l) \sim
(l_c-l)^\gamma$, close to $l_c$ with exponent
$\gamma^{\mathrm{MD}} \cong 4.2$. Our approach yields a power law
with $\gamma=1$, which significantly deviates from the MD result.
This discrepancy may be due to some uncertainties of the
simulation($\gamma^{\mathrm{MD}} \cong 4.2$ is an unusually high
value; for supercooled liquids $\gamma$ is close to two) or it may
originate from our approximations. With regard to the latter it
should be noted that the simplest version of Edwards and Vilgis'
description of the glass transition in a dense fluid of rods of
finite thickness\cite{EV} predicts a power law decay of the
self-diffusion coefficient with $\gamma=1$. This result is changed
in more complicated formulations of the theory. Whether our
approach can be developed in an analogous way is left for the
future study.

To summarize, we have considered
a system of hard rods fixed with their centers on
a simple cubic lattice. We approximated the influence of a third rod on
a pair of rods by introducing an effective diffusion tensor and
derived a closed set of equations for the 1- and 2-rod
density. These equations lead to a self consistent equation for a
rotational diffusion constant $D(l)$. In contrast to MCT
it is \textit{not} a static but a
\textit{dynamical} 2-particle correlator which determines the
slowing down of the dynamics and finally the glass transition.
Nevertheless, similar to the MCT, a cage effect exists in our model, 
too,
but the cage is of \textit{pure} dynamical nature.
If $l$ is large
enough, the tagged rod will perform librations within this
\textit{``dynamical cage''} built by all the neighbouring rods,
without existence of any static
correlations. Note that this mechanism exists even for rods with
\textit{finite} $d$, as long as the static correlations are
weak enough. With increasing $d$ there will be a crossover to the
``static'' cage effect as described by MCT. Whether or not MCT can
be extended such that this ``dynamical'' cage effect is included
is, of course, a challenge for the future. In any case, the
present model exhibits glassy behavior without possessing any
static correlations.
\bigskip

ACKNOWLEDGMENT
 \medskip

We are grateful to M. Fuchs for helpful comments and to M. Ricker
and D. Garanin for preparation of the figures. G. S. gratefully
acknowledges support from the National Science Foundation through
grant No. CHE 0111152 and the Alexander von Humboldt Foundation.

\newpage

\end{document}